\documentclass[useAMS,usenatbib]{mnras}
\usepackage{natbib}
\usepackage{color,graphicx,subcaption} 
\usepackage{amsmath}        
\usepackage{amsfonts}       
\usepackage{amssymb}        
\usepackage{wasysym}        
\usepackage{caption}

\title[MSP micro-structure with LEAP]{Detection of quasi-periodic micro-structure in three millisecond pulsars with the Large European Array for Pulsars}

\author[K.~Liu et al.]{K.~Liu,$^{1}$\thanks{kliu@mpifr-bonn.mpg.de} J.~Antoniadis$^{2,1,3}$, C. G. Bassa$^{4}$, S. Chen$^{5,6}$, I. Cognard$^{5,6}$, M.~Gaikwad$^{1}$, H. Hu$^{1}$, J.~Jang$^{1}$, \newauthor G. H. Janssen$^{4,7}$, R. Karuppusamy$^{1}$,  M. Kramer$^{1,8}$, K. J. Lee$^{9}$, R. A. Main$^{1}$, G. Mall$^{1,10,11}$, \newauthor J. W. McKee$^{10}$, M. B. Mickaliger$^{8}$, D. Perrodin$^{12}$, S. A. Sanidas$^{8}$, B. W. Stappers$^{8}$, L. Wang$^{8, 13}$, \newauthor W. W. Zhu$^{13}$, M. Burgay$^{12}$, R. Concu$^{12}$, A. Corongiu$^{12}$, A. Melis$^{12}$, M. Pilia$^{12}$, A. Possenti$^{12,14}$ \\
$^{1}$Max-Planck-Institut f\"{u}r Radioastronomie, Auf dem H\"{u}gel 69, 53121, Bonn, Germany\\
$^{2}$Institute of Astrophysics, FORTH, Dept. of Physics, University of Crete, Voutes, University Campus, GR-71003 Heraklion, Greece \\
$^{3}$Argelander Institut f\"{u}r Astronomie, Auf dem H\"{u}gel 71, 53121, Bonn, Germany\\
$^{4}$ASTRON, the Netherlands Institute for Radio Astronomy, Oude Hoogeveensedijk 4, 7991 PD Dwingeloo, The Netherlands \\
$^{5}$Laboratoire de Physique et Chimie de l'Environnement et de l'Espace LPC2E CNRS-Universit{\'e} d'Orl{\'e}ans, F-45071, Orl{\'e}ans, France \\
$^{6}$Station de radioastronomie de Nan{\c c}ay, Observatoire de Paris, PSL Research University, CNRS/INSU F-18330 Nan{\c c}ay, France \\
$^{7}$Department of Astrophysics/IMAPP, Radboud University, P.O. Box 9010, 6500 GL Nijmegen, The Netherlands \\
$^{8}$Jodrell Bank Centre for Astrophysics, School of Physics and Astronomy, The University of Manchester, Manchester M13 9PL,UK \\
$^{9}$Kavli institute for astronomy and astrophysics, Peking University, Beijing 100871,P.R.China \\
$^{10}$Canadian Institute for Theoretical Astrophysics, University of Toronto, 60 St. George Street, Toronto, ON M5S 3H8, Canada\\
$^{11}$Department of Physics, University of Toronto, 60 St. George Street, Toronto, ON M5S 1A7, Canada\\
$^{12}$INAF - Osservatorio Astronomico di Cagliari, via della Scienza 5, I-09047 Selargius (CA), Italy \\
$^{13}$National Astronomical Observatories, Chinese Academy of Sciences, A20 Datun Rd, Chaoyang District, Beijing 100012, P.\,R.\,China \\
$^{14}$Università di Cagliari, Dipartimento di Fisica, S.P. Monserrato-Sestu Km 0,700, I-09042 Monserrato (CA), Italy \\
}

\begin{document}
\maketitle
\begin{abstract}
We report on the detection of quasi-periodic micro-structure in three millisecond pulsars (MSPs), PSRs~J1022+1001, J2145$-$0750 and J1744$-$1134, using high time resolution data acquired with the Large European Array for Pulsars at a radio frequency of 1.4\,GHz. The occurrence rate of quasi-periodic micro-structure is consistent among pulses with different peak flux densities. Using an auto-correlation analysis, we measure the periodicity and width of the micro-structure in these three pulsars. The detected micro-structure from PSRs~J1022+1001 and J1744$-$1134 is often highly linearly polarised. In PSR~J1022+1001, the linear polarisation position angles of micro-structure pulses are in general flat with a small degree of variation. Using these results, we further examine the frequency and rotational period dependency of micro-structure properties established in previous work, along with the angular beaming and temporal modulation models that explains the appearance of micro-structure. We also discuss a possible link of micro-structure to the properties of some of the recently discovered fast radio bursts which exhibit a very similar emission morphology. 
\end{abstract}


\section{Introduction} \label{sec:intro}
Pulsars exhibit a large variety of radio emission phenomena on a wide range of timescales, which is still to be fully understood. The very first pulsar observations already revealed the significant variability among the pulses from every rotational period. In particular, some pulses show concentrated emission in sub-millisecond features, usually with a typical width and sometimes a quasi-periodicity \citep{ccd68,han72}. These so-called \textit{micro-structure} phenomena have been seen in a number of canonical pulsars \citep[e.g.,][]{cwh90,lkwj98,kjv02}, and recently also in millisecond pulsars \citep[MSP;][]{dgs16}. Many of the micro-structure pulses exhibit significant fraction of polarisation \citep[e.g.,][]{mar15,dgs16}, higher than those expected from the average pulse profile. Simultaneous observations at multiple frequencies have shown that micro-structures occur over a wide frequency range at the same time \cite[e.g.,][]{rhc75,bfs81}, which suggests a fundamental association of micro-structure with the pulsar emission process. 

So far, a number of models have been developed to explain the appearance of micro-structure in pulsars. Generally, these models can be categorized into two types of scenarios. The first one involves an angular radiation pattern as a result of multiple thin flux tubes along the magnetic field lines where streaming bunches of charged particles radiate in the direction of propagation \citep[e.g.,][]{ben77}. As the pulsar rotates, the emission structure is then sampled in time. The width of the micro-structure, in this case, would correspond to the angular beam width of the radiation. The second scenario includes a temporal intensity modulation of the emission, and may be caused by electrodynamical fluctuations \citep{cr80,bb80}, neutron star vibrations \citep{van80,cr04b} or radiation transfer effects \citep{ht81,cjd04}. Here, the micro-structure width reflects the actual timescale of the emission, which may be related to the radial structure in the plasma outflow.

Most of micro-structure studies focus on measuring their temporal width and quasi-periodicity, and relating them with other properties such as pulsar rotational period and observing frequency \citep[e.g.,][]{cor79,kjv02,mar15}. These help to distinguish between the two possible scenarios for micro-structure, possibly identifying the physical mechanism of the emission. To date, the majority of micro-structure investigations have been based on canonical pulsars. Extending the analysis to MSPs would allow to examine the existing theories in a broader parameter (e.g., the pulsar period) space and with new samples. Comparing the results obtained from canonical pulsars and MSPs would also reveal whether micro-structure is a common feature for all pulsar populations. 


Finding micro-structure emission in MSPs is in general difficult mainly due to the limitation of sensitivity and other technical constraints such as time and frequency resolution of the recorded data. As a result, the first micro-structure detection in MSPs has only been achieved recently \citep{dgs16}, following a small number of reported non-detections \citep{jak+98,sal98,jap01,lbj+16}. The Large European Array for Pulsars (LEAP) is the ideal instrument to study micro-structure emission from pulsars. LEAP coherently combines pulsar observations from the five largest radio telescopes in Europe: the 100-m Effelsberg Telescope in Germany, the 76-m Lovell Telescope at Jodrell Bank Observatory (JBO) in the U.K., the 94-m equivalent Nan\c{c}ay Radio Telescope in France, the 64-m Sardinia Radio Telescope in Italy, and the 94-m equivalent Westerbork Synthesis Telescope in the Netherlands. This delivers a sensitivity equivalent to a 194-m single dish \citep{bjk+16}. The boost in sensitivity is not only important for high-precision pulsar timing, but has also allowed for other pulsar projects that are largely limited by sensitivity with smaller telescopes, such as single-pulse and scintillation studies of MSPs \citep{lbj+16,msb+19,msa+20}. In particular, the baseband recording and storage capability enables LEAP to generate data products with customised time and frequency resolutions and coherent de-dispersion for these studies.

The rest of the paper is organized as follows. In Section~\ref{sec:obs} we describe the observations and post-processing of the data. In Section~\ref{sec:res} we present the detection of micro-structure in three MSPs and the measurements of their properties. Section~\ref{sec:dis} includes a brief discussion on the results and conclusions can be found in Section~\ref{sec:conclu}.

\section{Observations} \label{sec:obs}
\begin{figure*}
\centering
\hspace*{-1.6cm}
\vspace*{-0.3cm}
\includegraphics[scale=0.4]{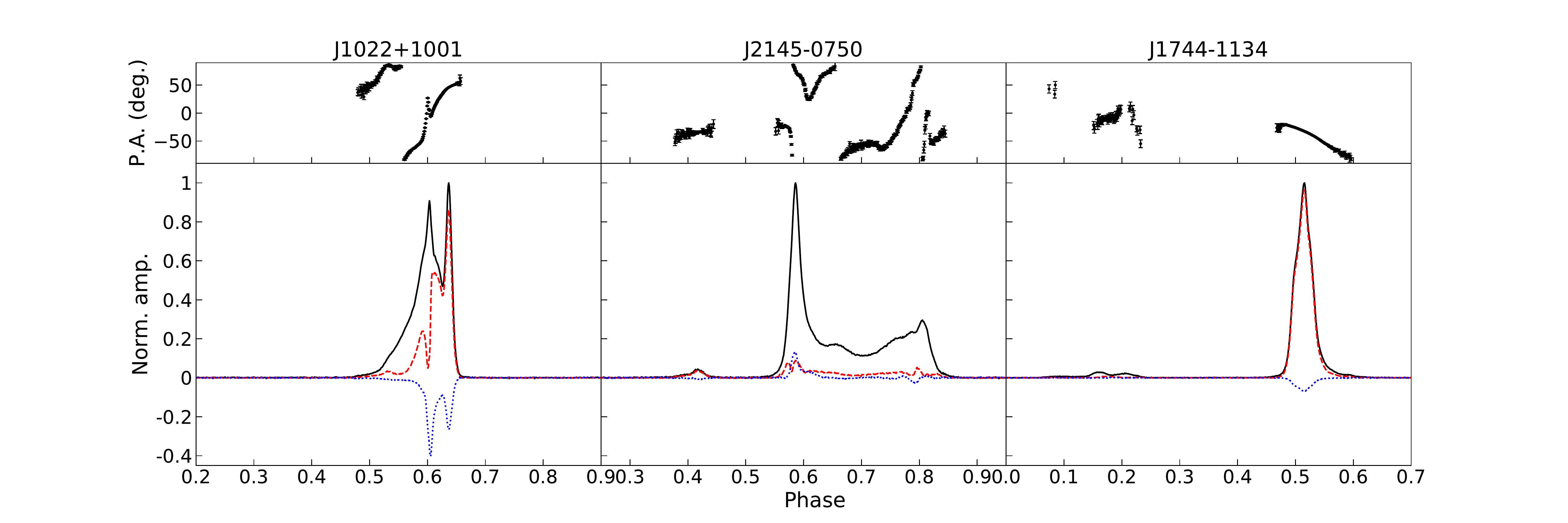}
\caption{Polarisation profiles and the linear polarisation position angles (P.A.) of the three MSPs from our observations listed in Table~\ref{tab:obs}. The black solid, red dashed and blue dotted lines represent the total intensity, linear and circular polarisation, respectively. \label{fig:polprofs}}
\end{figure*}

LEAP conducts monthly observations of over twenty MSPs at L-band (1396\,MHz), for the main purpose of detecting low-frequency gravitational waves with the PTA experiment \citep[][]{ccg+21}. To allow studies of micro-structure in MSPs, observations where a significant number of single pulses can be clearly detected, are needed. Therefore, here we selected three bright MSPs, J1022+1001, J2145$-$0750, J1744$-$1134. For each source, we selected the observation epoch when the pulsar signal was greatly amplified by interstellar scintillation, achieving the highest signal-to-noise ratio (S/N) of detection per unit time in the set of all our observations. During the observation, the astronomical signals were sampled at Nyquist rate with 8 bits and recorded on spinning disks as baseband voltages with a bandwidth of 128\,MHz. The data collected at each individual telescope were later transferred to our storage server at JBO, where they were correlated, calibrated for polarisation, and coherently combined using a dedicated software correlator \citep{sbj+17}. Next, the combined baseband data from those epochs were processed to produce single-pulse data (i.e., one archive per rotation) with the specifications given in Table~\ref{tab:obs}. The single-pulse data were then cleaned to remove narrow band radio interference. The rotation measure values from \cite{dhm+15} were used to correct for the Faraday rotation in the polarisation. These data-processing steps used the \textsc{psrchive} software package \citep{hsm04}. Figure~\ref{fig:polprofs} presents the average profiles of the three pulsars each obtained from the single-epoch observation listed in Table~\ref{tab:obs}. They are are highly consistent with previously reported averaged profiles obtained by adding many hours of observations \citep[e.g.,][]{dhm+15}, demonstrating the high quality of the coherent combined LEAP data. 

\begin{table*}
\centering
\begin{center}
\caption[]{Properties of the pulsars and details of the data investigated in this paper. Here $P$, $T_{\rm int}$, $\Delta t$, $N_{\rm p}$ and $\rho_{\rm S/N>6}$ represent the pulsar rotational period, duration of the observation, time resolution of the single-pulse data, total number of pulses recorded and the percentage of pulses with peak S/N higher than 6. The single-pulse data for each pulsar were coherently de-dispersed with the DM values given below and retain a 1-MHz frequency resolution over the 128-MHz bandwidth. The measurements of micro-structure (median) quasi-periodicity ($P_{\rm \mu}$) and width ($\tau_{\rm \mu}$) are also listed. }
\begin{tabular}{cccccccccc}
\hline
Jname &$P$ (ms) & DM (cm$^{-3}$\,pc) &MJD & $T_{\rm int}$ (min) &$\Delta t$ ($\mu$s) &$N_{\rm p}$ &$\rho_{\rm S/N>6}$ &$P_{\rm \mu}$ ($\mu$s) & $\tau_{\rm \mu}$ ($\mu$s)\\
\hline
J1022+1001 & 16.45 &10.2595 & 56739.9 & 43.0 & 2.03 &$\sim154,500$ &7.4\% &$14.9^{+5.2}_{-2.8}$ & $9\pm 1$\\
J2145$-$0750 & 16.05 &8.9953 & 56740.4 & 32.0 & 1.96 &$\sim119,600$ & 4.2\% &$17.8^{+2.7}_{-3.2}$ & $10\pm 1$\\
J1744$-$1134  & 4.07 &3.1380 & 57107.2 &38.8 & 0.99 &$\sim567,100$ &1.4\% &$6.0^{+2.4}_{-0.8}$ & $3.9\pm 0.5$\\
\hline
\end{tabular}
\label{tab:obs}
\end{center}
\end{table*}

\section{Results} \label{sec:res}
The single-pulse data were used to search for and investigate micro-structure emission from each of the selected pulsars. We first calculated the peak S/N of all single pulses and selected those with $\mathrm{S/N}>6$. This left 7.4, 4.2, 1.4 percent of all recorded pulses for PSR~J1022+1001, J2145$-$0750, J1744$-$1134, respectively. We then calculated the auto-correlation function (ACF) of the on-pulse region for each of the selected pulses. The presence of quasi-periodic micro-structure would manifest itself by exhibiting equally spaced maxima in the ACF, with the time lag of the first maxima corresponding to the characteristic temporal separation, i.e., periodicity of the micro-pulses \citep[e.g.,][]{cor79,lkwj98,kjv02}. We also generated the ACF with a linear fit (to itself) subtracted, which in some cases makes the maxima in the ACF more prominent. Next we visually checked all the selected pulses along with their ACFs to identify those exhibiting micro-structure emission. For pulses with regularly spaced maxima in the ACF, we recorded the time lag of the first maxima as the quasi-periodicity $P_{\rm \mu}$ of the micro-structure. The micro-structure width $\tau_{\rm \mu}$ was measured using the average ACF of all pulses with micro-structure detected, which corresponds to the first sign change in the ACF slope \citep[see e.g., Figure 1 of ][]{kjv02}.

\begin{figure*}
\centering
\includegraphics[scale=0.34]{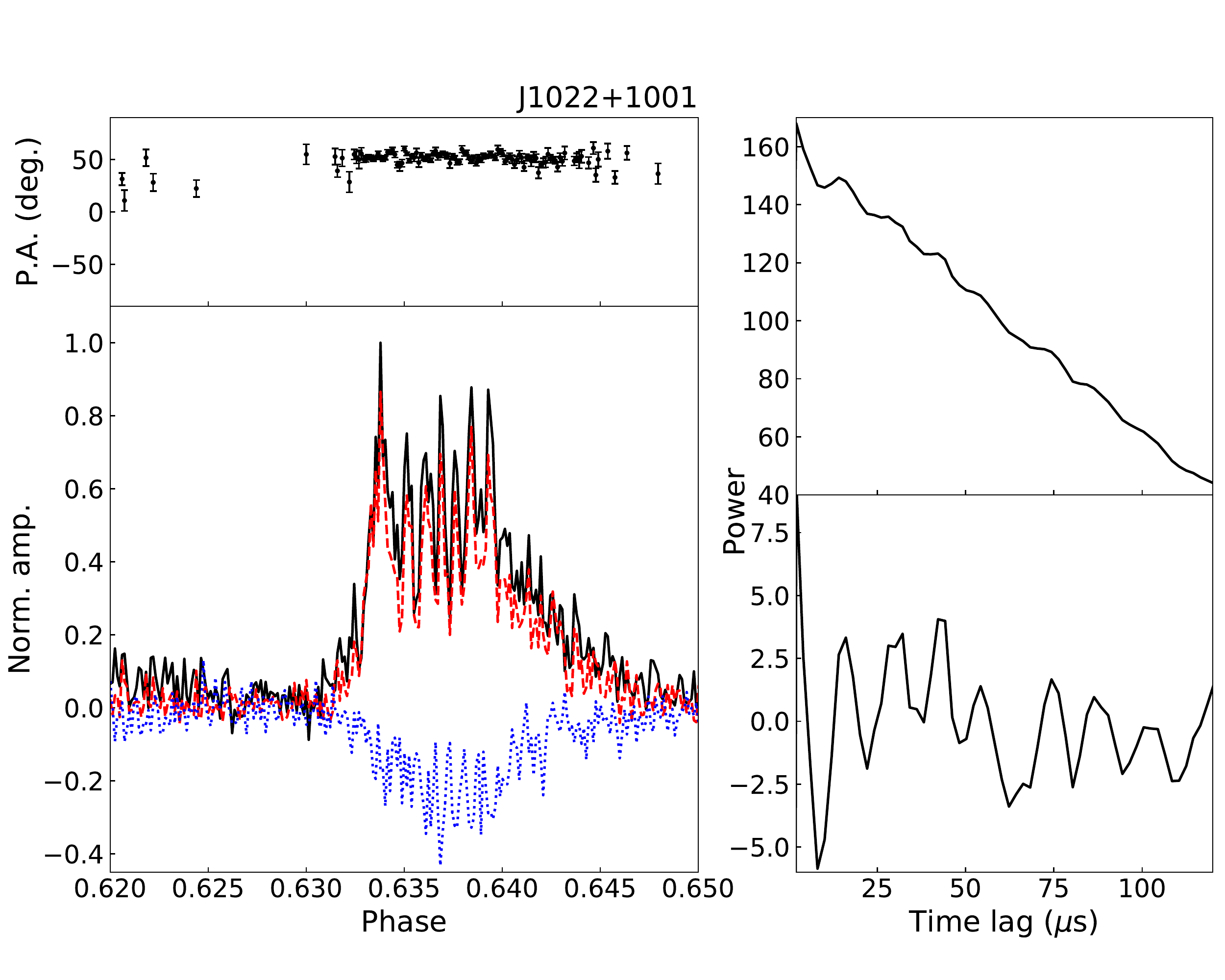}
\includegraphics[scale=0.34]{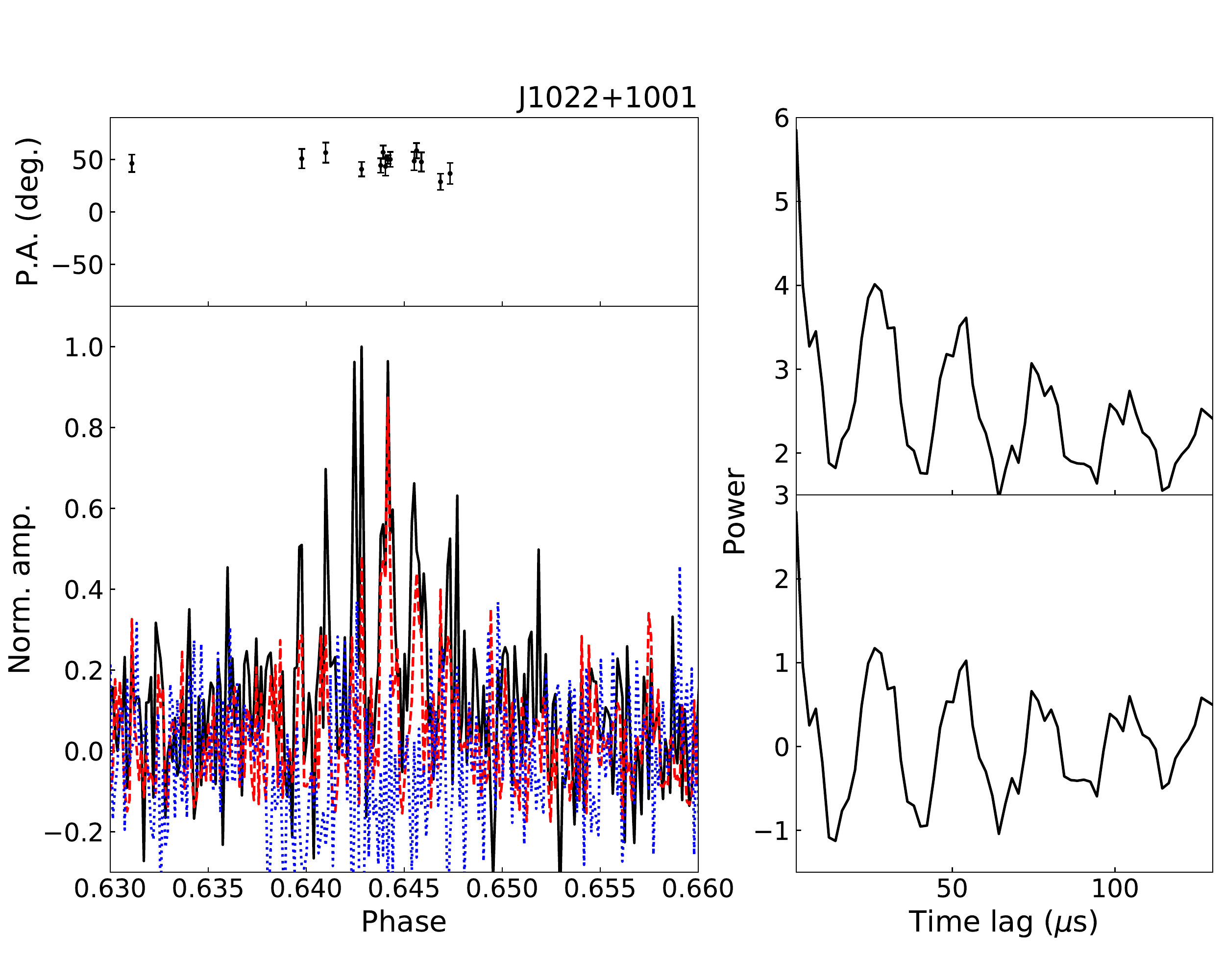}
\includegraphics[scale=0.34]{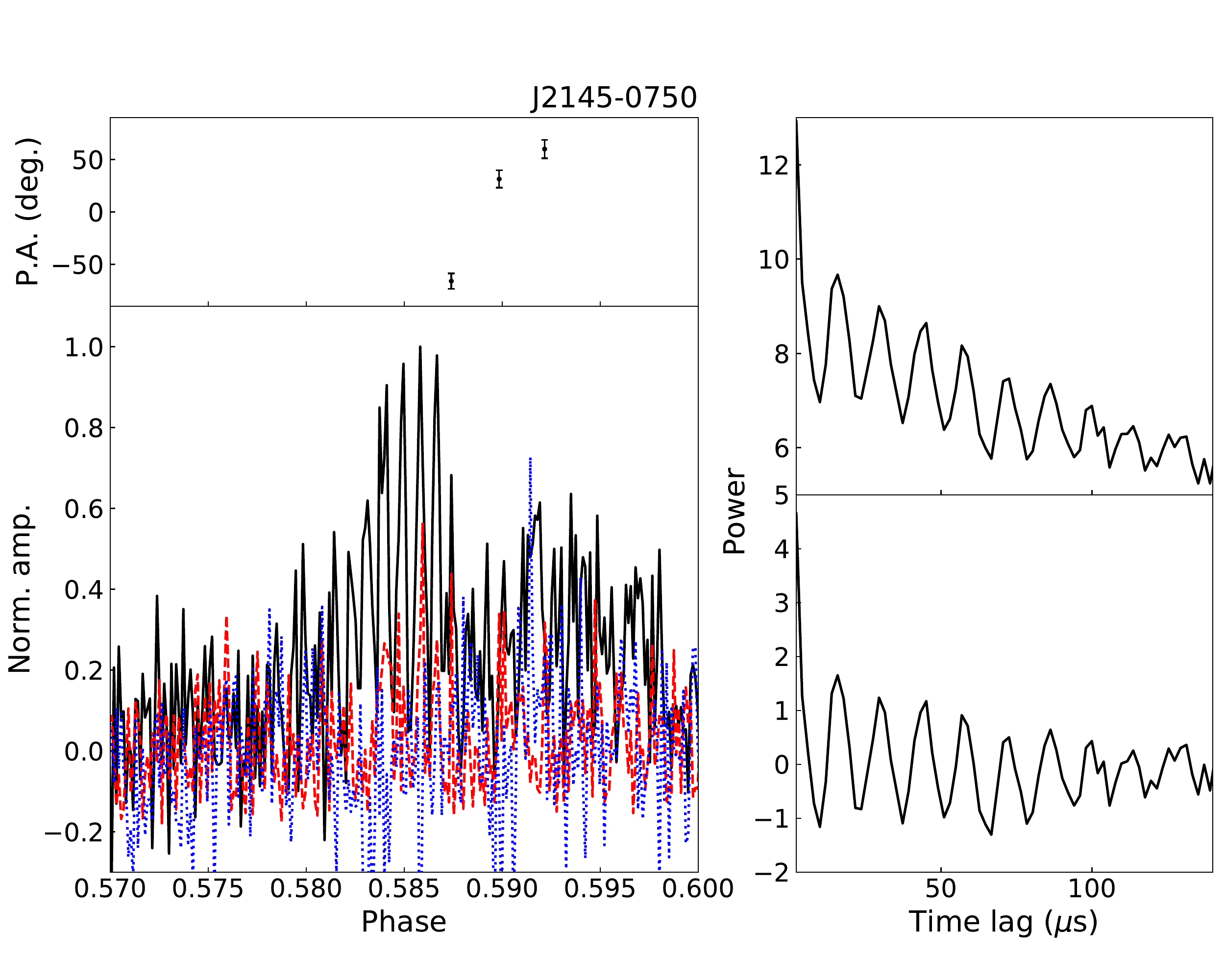}
\includegraphics[scale=0.34]{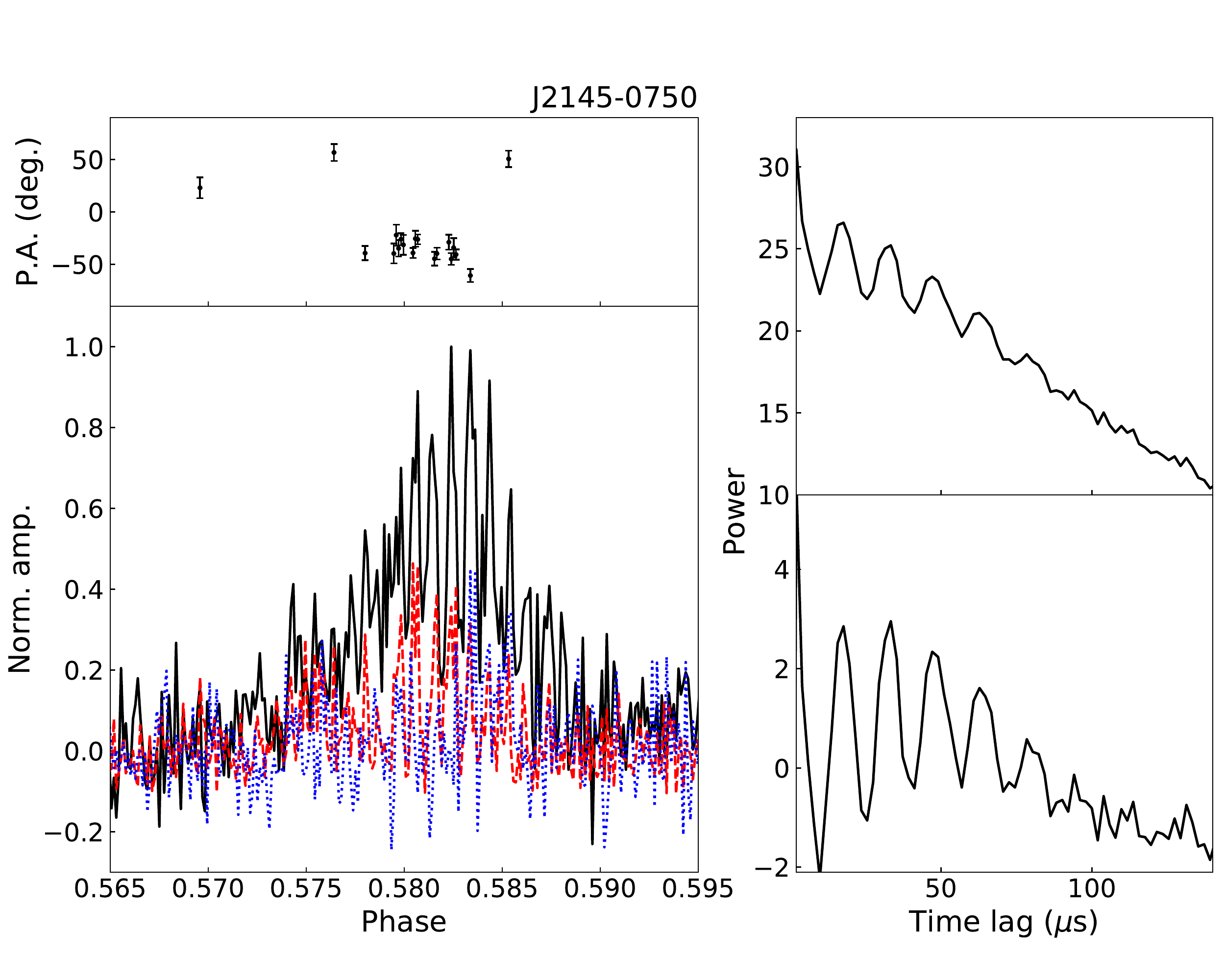}
\includegraphics[scale=0.34]{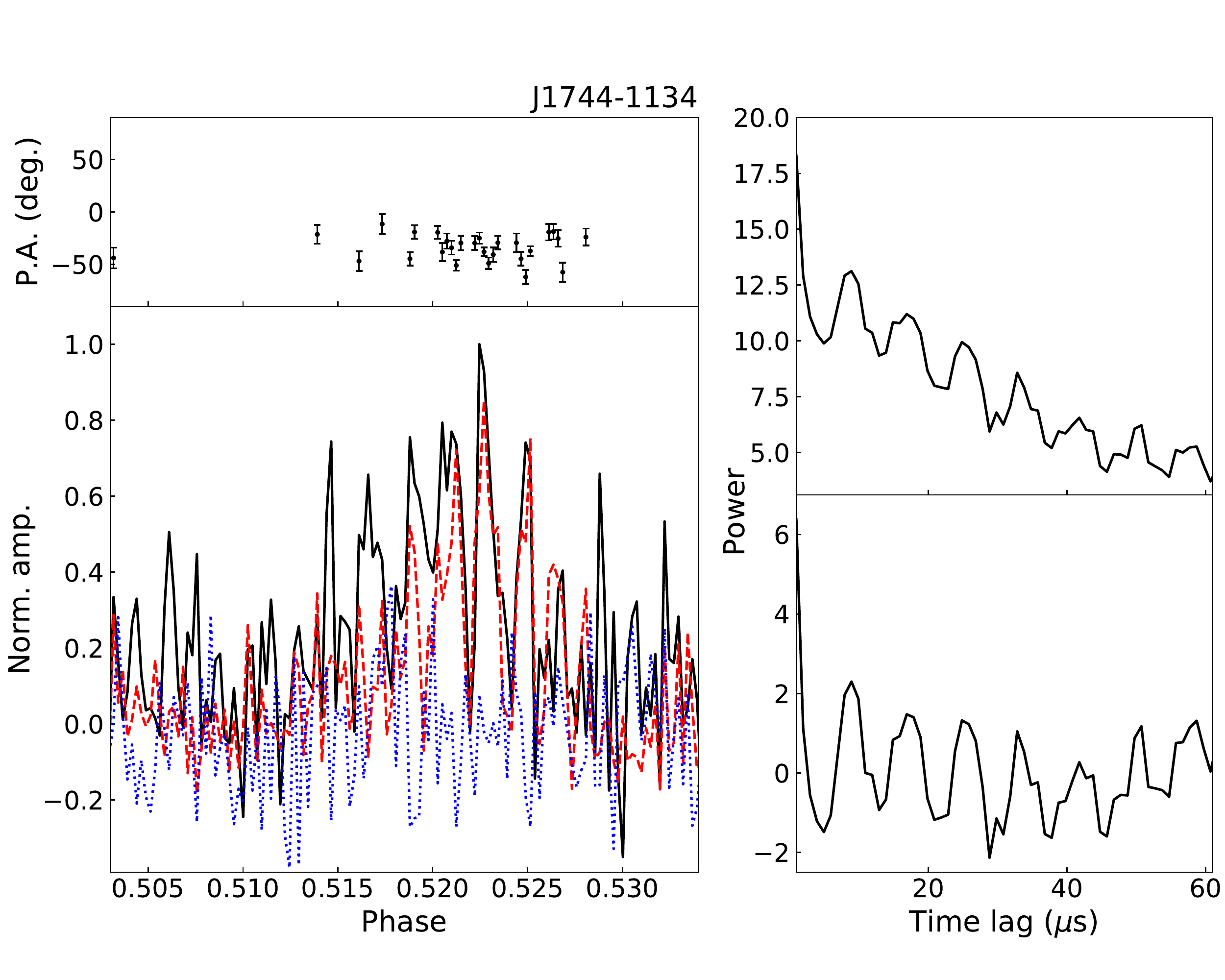}
\includegraphics[scale=0.34]{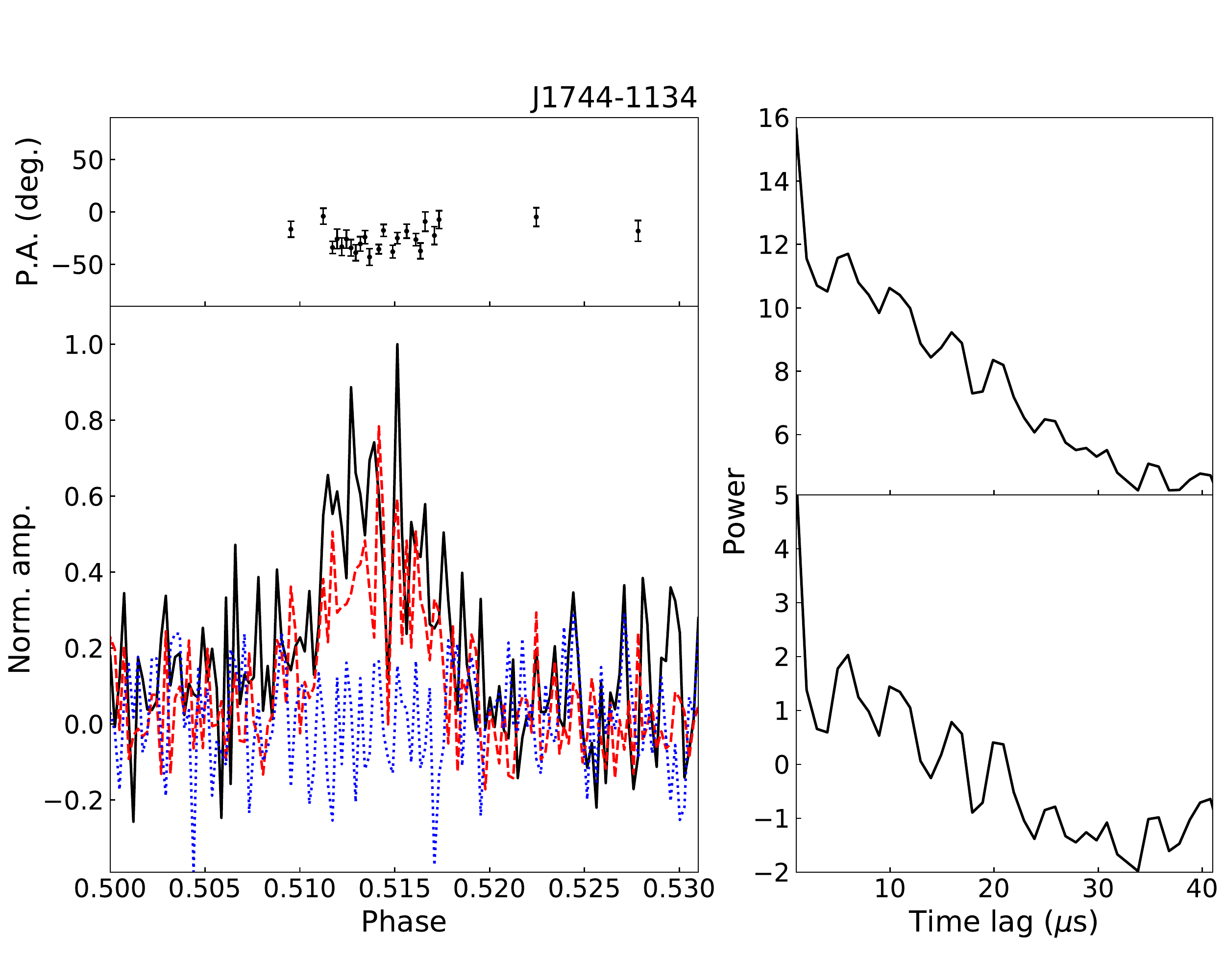}
\caption{Examples of pulses showing quasi-periodic micro-structure from PSRs J1022+1001 (first row), J2145$-$0750 (second row) and J1744$-$1134 (third row). The first and third column of panels show the pulse intensities and their linear polarisation position angles (P.A.), where the black solid, red dashed and blue dotted lines represent total intensity, linear and circular polarisation, respectively. The second and forth column of panels present the ACFs of the pulses shown on their left (upper), and the same ACFs but with a linear fit subtracted (lower). Note that for each pulsar, the two example pulses are aligned in phase with respect to the rotational period. 
\label{fig:micropulses}}
\end{figure*}

\subsection{J1022+1001} \label{ssec:1022}
\begin{figure*}
\centering
\includegraphics[scale=0.36]{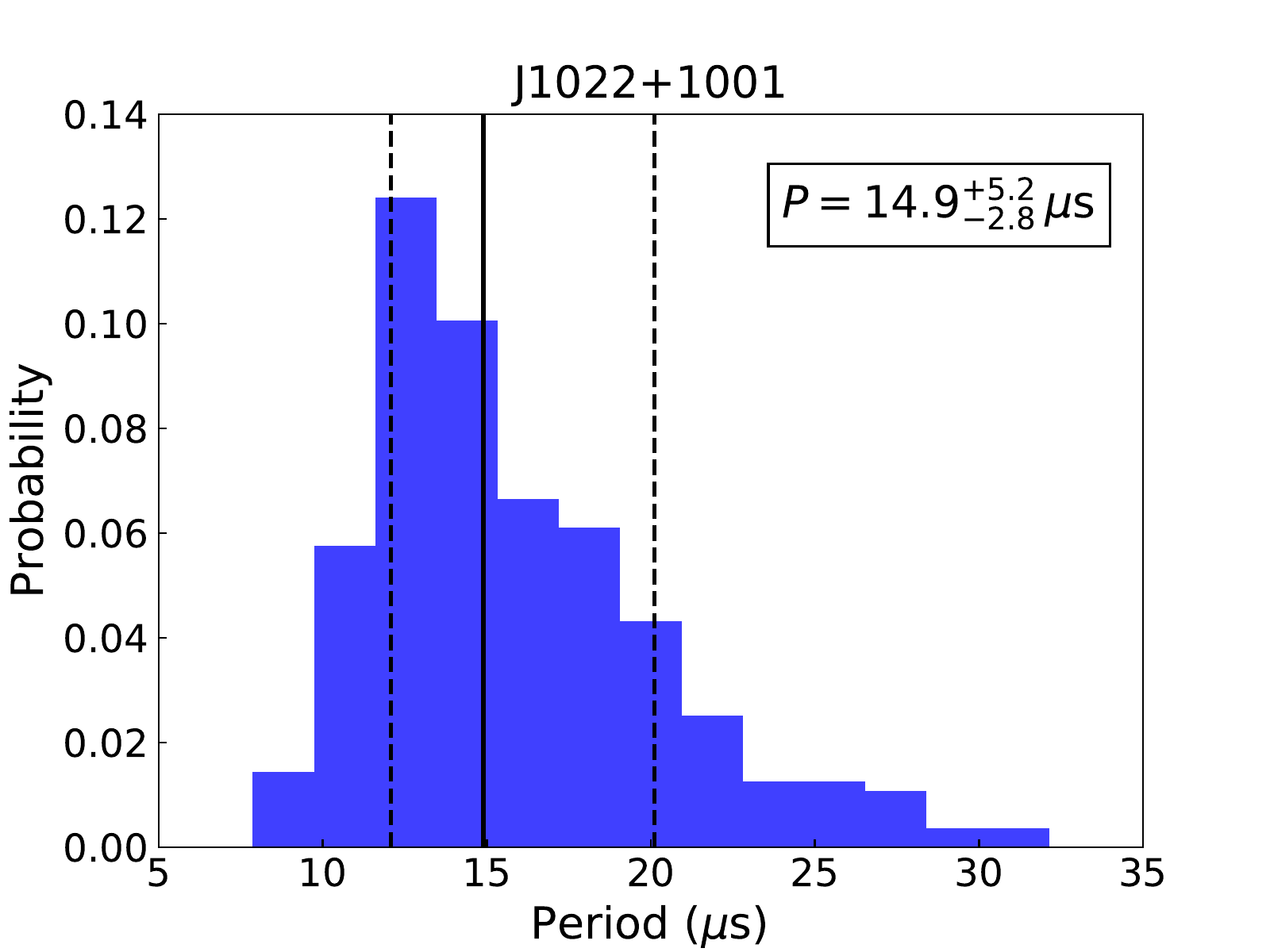}
\includegraphics[scale=0.36]{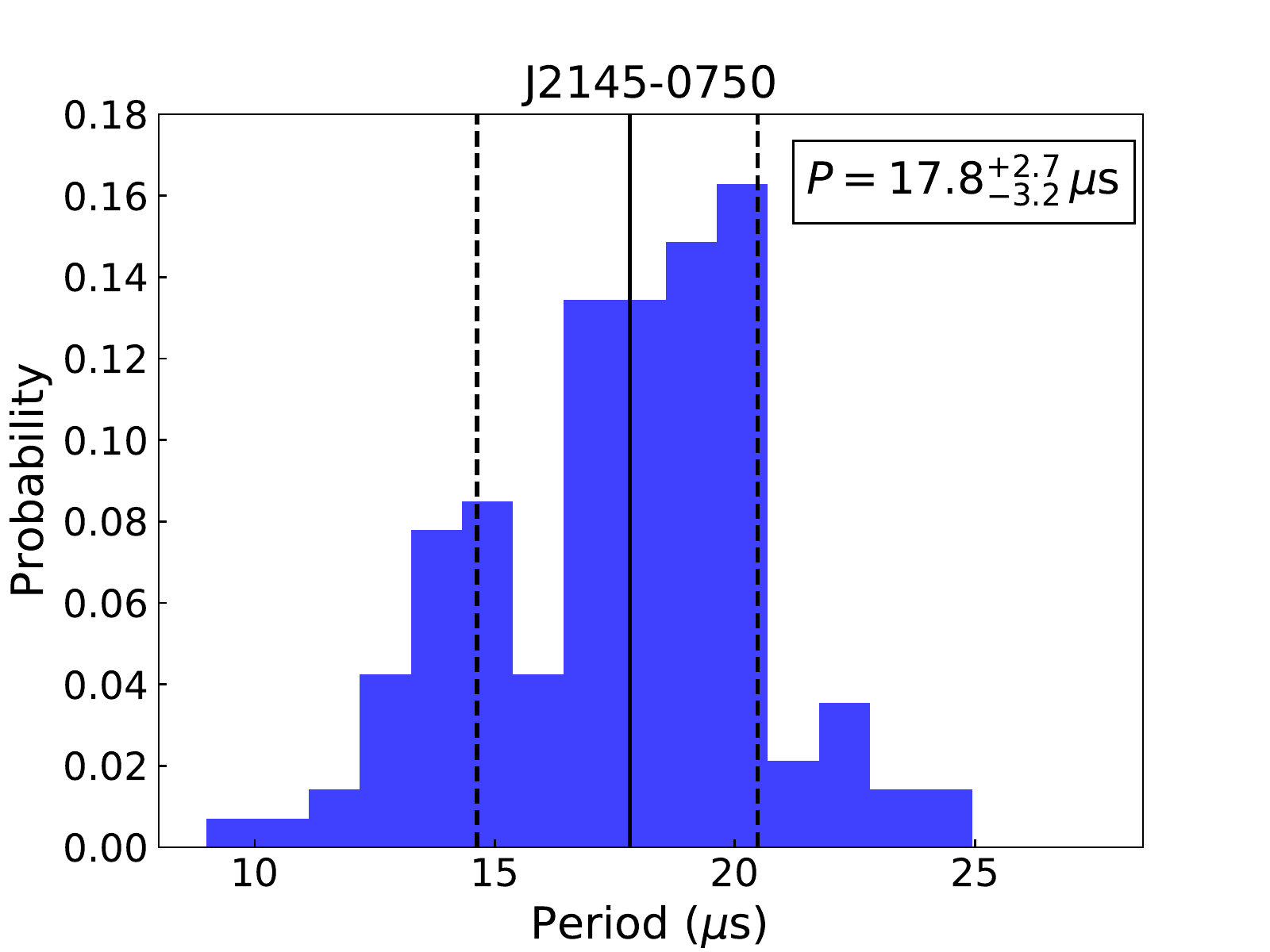}
\includegraphics[scale=0.36]{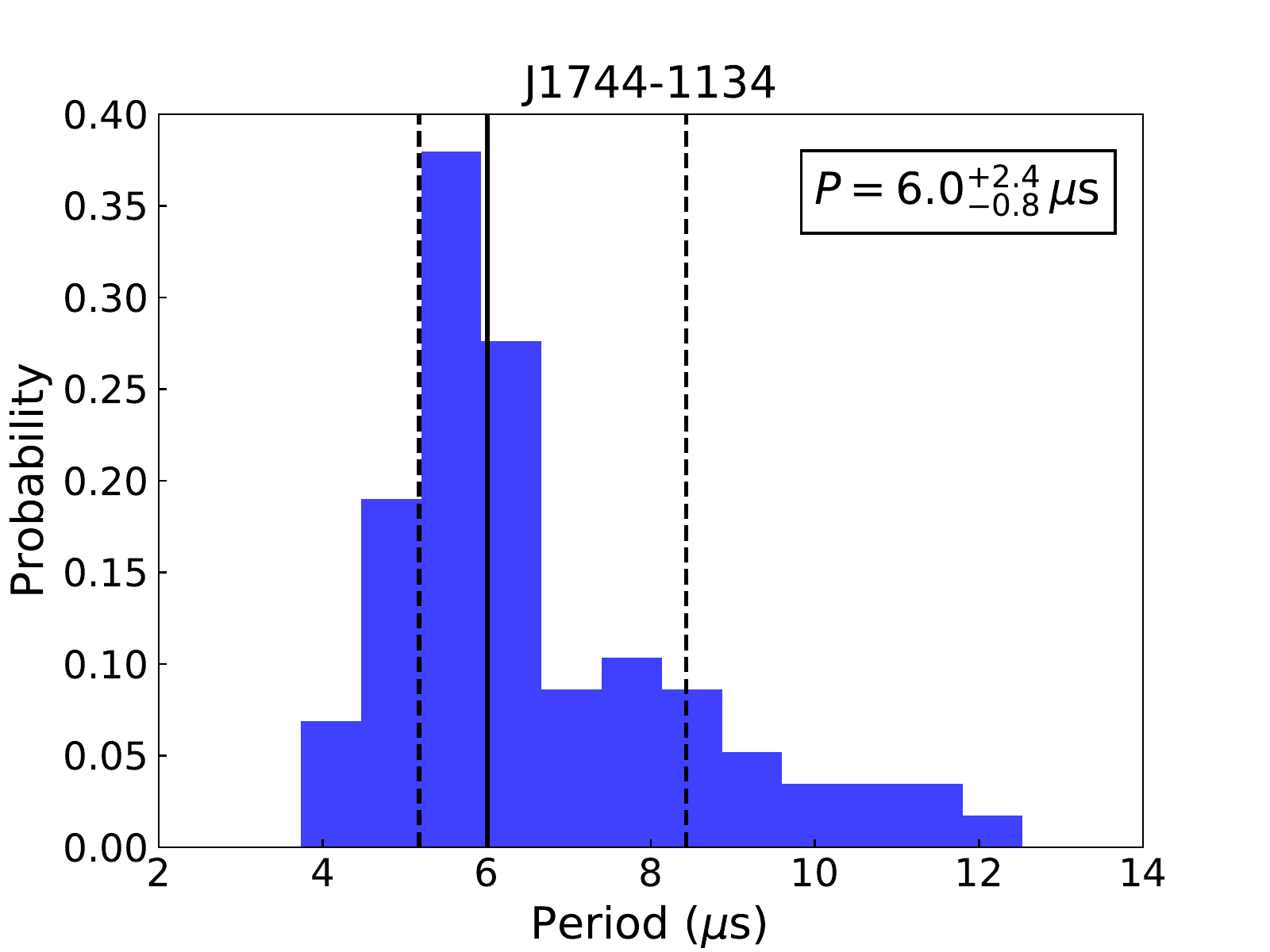}
\caption{Histogram of micro-structure periodicity measured in J1022+1001 (left), J2145$-$0750 (middle) and J1744$-$1134 (right). In each plot, the solid line represents the median of the distribution and the two dashed lines marks the 1-$\sigma$ confidence interval counting from the median to both sides. \label{fig:phist}}
\end{figure*}

\begin{figure*}
\centering
\includegraphics[scale=0.38]{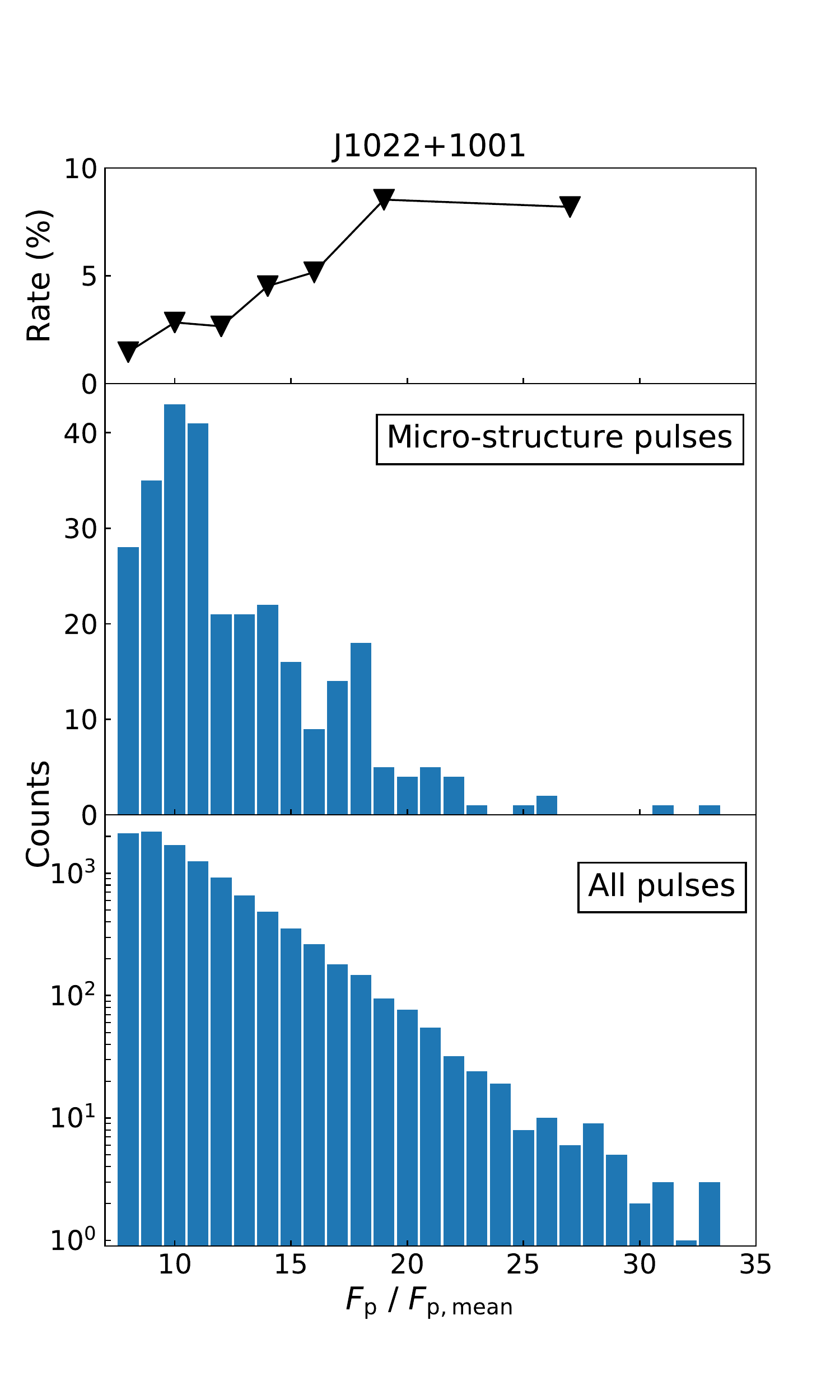}
\includegraphics[scale=0.38]{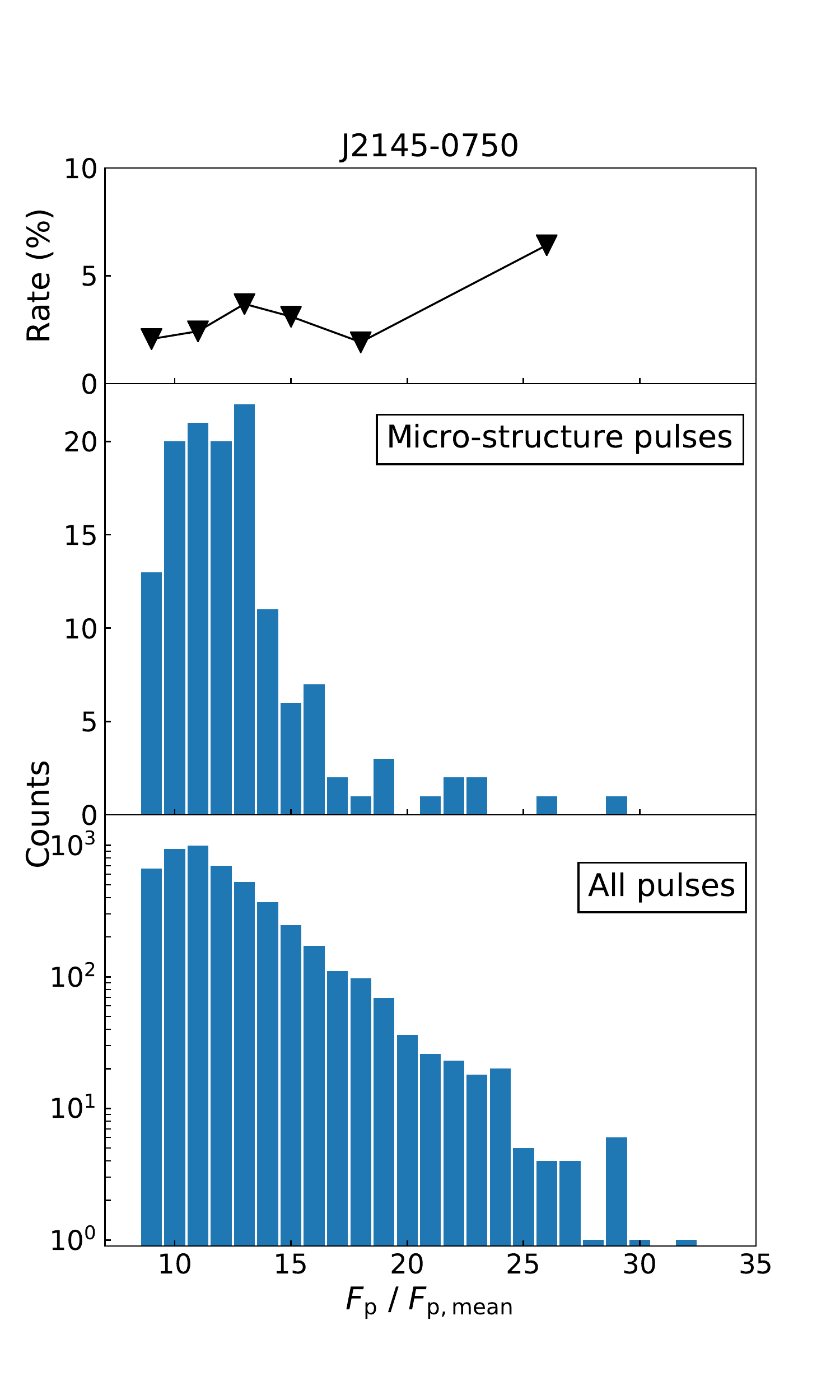}
\includegraphics[scale=0.38]{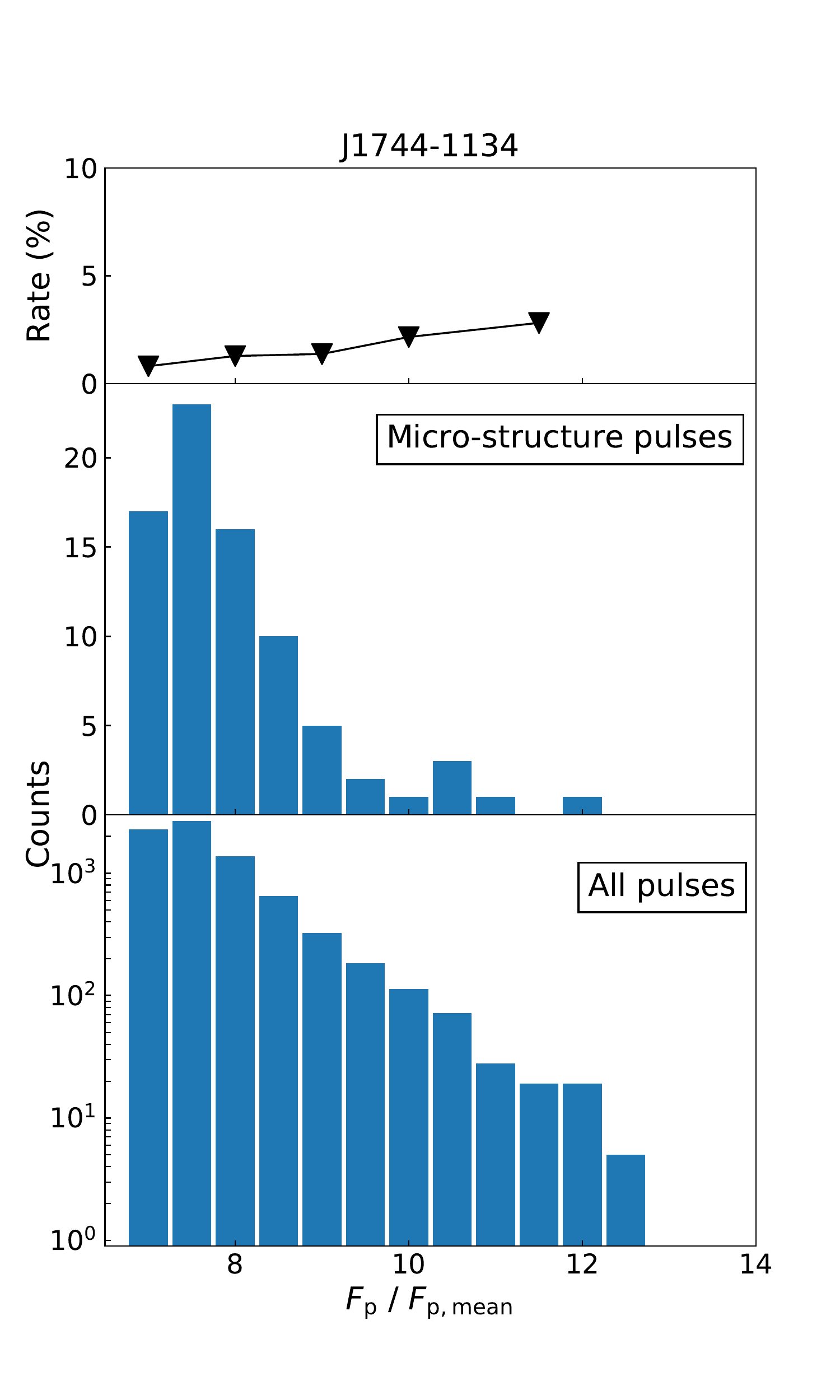}
\caption{Distribution of normalised peak flux density (with respect to the mean of all pulses) of micro-structure pulses in comparison with that of all pulses, for J1022+1001 (left), J2145$-$0750 (middle) and J1744$-$1134 (right). The top panels show the percentage detection rate of quasi-periodic micro-structure pulses. \label{fig:powerdist}} \end{figure*}

Approximately 37\% of the S/N$>$6 pulses that have been viewed show micro-structure emission, and 3\% exhibit quasi-periodicity. The top panels in Figure~\ref{fig:micropulses} show two typical examples. The vast majority of these pulses coincide in phase with the trailing component of the average pulse profile (see Figure~\ref{fig:polprofs}). This is consistent with the findings in \cite{lkl+15} where most of the detected bright single pulses are from the trailing component as a result of its high intensity modulation \citep{es03b}. Most micro-structure pulses are seen to have a high degree of linear polarisation and some with a clear circular component, in agreement with the average polarisation profile. Figure~\ref{fig:phist} presents the distribution of the measured periodicities for those showing quasi-periodic features, which ranges in the 10--30\,$\mu$s interval with a median $P_{\rm \mu}$ of 14.9\,$\mu$s. In addition to the total intensity, we also searched for quasi-periodic micro-structure in the linear and circular components of the pulses and found consistent periodicities, similar to what has been reported by \cite{mar15} for canonical pulsars. Averaging the ACFs from all these pulses, we obtained a characteristic width of $\tau_{\mu}=9\pm 1$\,$\mu$s. Figure~\ref{fig:powerdist} shows the distribution of quasi-periodic micro-structure pulses with respect to their peak flux density relative to the average of all pulses. It can be seen that quasi-periodic micro-structure occurs in pulses of different peak flux densities, including the brightest ones. The detection rate is found to be roughly independent of the peak flux density, with a potential small preference to stronger pulsars, which however, needs to be confirmed with more samples.

\begin{figure}
\centering
\includegraphics[scale=0.58]{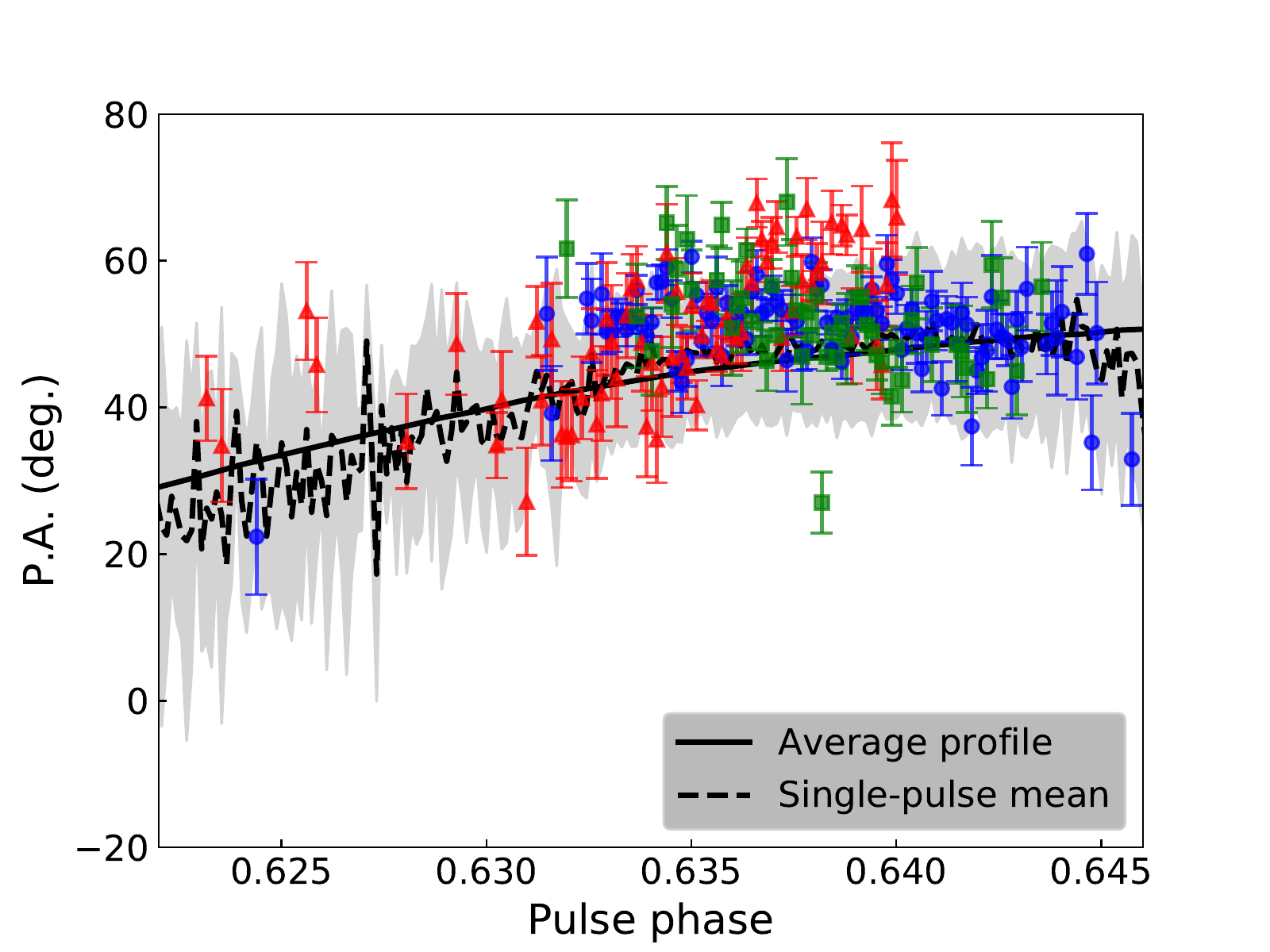}
\caption{Comparison of P.A. swings from PSR~J1022+1001. The solid black line is the P.A. swing from the integrated profile shown in Figure~\ref{fig:polprofs}. The dashed black line represents the weighted mean from all micro-structure pulses, while the shallow grey band represents the 1-sigma probability range. The red triangles, blue dots, and green squares are P.A. values from three example micro-structure pulses, respectively. The first shows a steeper increase of P.A. in pulse phase, the second is mostly steady and the third has an opposite sign of gradient. \label{fig:1022PA}}
\end{figure}

Figure~\ref{fig:1022PA} compares the linear polarisation position angles (P.A.) of the quasi-periodic micro-structure pulses with that obtained from the average profile. Here for each individual phase bin, a probability density distribution of P.A. was formed based on the measurements from these pulses. It can be seen that on average, the P.A. swings from the micro-structure pulses and average profile are highly consistent. Still, the P.A. swing from each pulse does exhibit additional variations within a small range. This can be seen in the three examples shown in Figure~\ref{fig:1022PA}, which exhibit different gradients of change in pulse phase. Overall, the P.A. values of micro-structure pulses fall in a narrow range, mostly because they are obtained from a narrow window in pulse phase, which corresponds to approximately only 1.5\% of the entire rotational period. This is significantly narrower compared to the entire on-pulse region of the average profile, which spans 20\% of the spin period as shown in Figure~\ref{fig:polprofs}.

\subsection{J2145$-$0750} \label{sssec:2145}
Approximately 38\% of the S/N$>$6 pulses show micro-structure and 3\% exhibit quasi-periodicity. Two examples of quasi-periodic micro-structure can be found in Figure~\ref{fig:micropulses}. All micro-structure detections cluster around the phase of the primary component in the averaged pulse profile (see Figure~\ref{fig:polprofs}). The micro-structures generally are weakly polarised, which is in line with the average polarisation profile shown in Figure~\ref{fig:polprofs}, and seemingly different from the finding at 630\,MHz reported by \cite{dgs16}. Still, occasionally some of the micro-pulses do show a high degree of polarisation which is up to $\sim$100\% and can be either linear or circular. As shown in Figure~\ref{fig:phist}, the measured micro-structure periodicities range in the 10--25\,$\mu$s interval with a median $P_{\rm \mu}$ of 17.8\,$\mu$s. From the number distribution in Figure~\ref{fig:powerdist}, it can be seen that the occurrence rate of micro-structure is in general uniform as a function of peak flux density, including the brightest group of pulses. The micro-structure timescale was measured to be $\tau_{\mu}=10\pm 1$\,$\mu$s from the averaged ACFs of all micro-structure pulses.

\subsection{J1744$-$1134}
About 14\% of the pulses that have been visually inspected show micro-structure, and 1\% exhibit quasi-periodic features. Two examples of the quasi-periodic micro-structure detections are found in Figure~\ref{fig:micropulses}. Similar to the situation in J2145$-$0750, all of these detections are associated with the primary component in the average profile shown in Figure~\ref{fig:polprofs}. As seen in Figure~\ref{fig:phist}, the measured periodicity of the micro-structures varies from 4 to 12\,$\mu$s, with a median $P_{\rm \mu}$ of 6.0\,$\mu$s. Averaging the ACFs of all micro-structure pulses, we obtained a micro-structure timescale of $\tau_{\mu}=3.4\pm 0.5$\,$\mu$s, after correcting for the sampling width. Most of the micro-structure pulses exhibit a high degree of linear polarisation, in line with the feature seen in the average profile as shown in Figure~\ref{fig:polprofs}. The linear component also shows periodicities consistent with those from the total intensity. As shown in Figure~\ref{fig:powerdist}, the occurrence rate of quasi-periodic micro-structure is generally constant for pulses of different peak flux densities. 

\section{Discussions} \label{sec:dis}

\subsection{Detection rate of micro-structure}
The fractions of pulses detected with quasi-periodic micro-structure are small in comparison with those reported in many canonical pulsars \citep[e.g.,][]{lkwj98}. In theory, an apparent quasi-periodicity may be produced by an uncorrelated intensity variation at different pulse phases, given that the number of samples is significant enough. To examine the potential impact of such an effect on our results, for each pulsar we randomly chose a stack of 1500 pulses and carried out the following experiment. For each individual pulse phase, we first randomly shuffled the order of intensity bins in the stack. We then calculated the ACFs of the pulses in the new stack and visually inspected the result to obtain the number of detections. This experiment has been carried out to all of the three pulsars, which gave detection rates of quasi-periodic micro-structure pulses of 0.4\%, 0.3\%, 0.1\% for PSRs J1022+1001, J2145$-$0750, J1744$-$1134, respectively. These are all approximately an order of magnitude lower than the those from the real data. This suggests that the detected quasi-periodic micro-structures in these pulsars, are unlikely to be solely attributed to uncorrelated intensity variation in pulse phase.

\subsection{Frequency dependency of micro-structure}
The micro-structure periodicity of PSR~J2145$-$0750 measured with our L-band observation is highly consistent with the value obtained at 610\,MHz by \cite{dgs16}, suggesting that the micro-pulse separation of this pulsar is likely to be frequency independent. This is similar to what has been observed so far in many young pulsars \citep{cwh90,mar15}, but is the first time it has been observed in a MSP. The frequency independence of micro-structure separation better supports the temporal radial origin of micro-structure scenario \citep{cwh90}, as in the angular beaming model, the separation is supposed to evolve as a function of observing frequency if it follows the radius-to-frequency mapping. However, we noticed that the pulse profile of J2145$-$0750 has a very similar width within a wide frequency range from 100\,MHz to at least 5\,GHz \citep{kll+99}. Though the profile shape of PSR~J2145$-$0750, primarily the amplitude ratio between the two main components, changes significantly in frequency, this may well be explained by the difference in spectral index of emission components as seen in other pulsars \citep[e.g.,][]{kjl+11}. Thus, these suggest that the frequency dependency of micro-structure is unlikely to be significant in J2145$-$0750 due to the fairly consistent emission beam width at multiple frequencies.

Our 3\% detection rate of micro-structure in J2145$-$0750 is a factor of a few lower than that reported by \cite{dgs16}. While the observing times investigated are approximately the same, \cite{dgs16} reported over 700 pulses with an S/N above 15 but there are only 33 in our case. To obtain a total number of 700 pulses from our observations, the S/N threshold needs to be set at 9.3. The detection rate for this group is approximately 3.5\% and thus with no significant difference. The lower rate of quasi-periodic micro-structure at L-band may be a result of a steeper spectrum of the micro-pulses or a lower pulse intensity modulation at higher frequencies as seen in PSR~B2016+28 \citep{cwh90}.


\subsection{Rotational period dependency of micro-structure}

The dependency on pulsar rotational period of the micro-structure periodicity and width, have been established in canonical pulsars for decades \citep{cor79,kjv02}, but only recently have they been investigated in MSPs \citep{dgs16}. These relations could provide important indication of the mechanism behind micro-structure, thus constraining the angular and temporal models that explain the appearance of quasi-periodic micro-structure \citep[e.g.,][]{cwh90}. Using a number of measurements in canonical pulsars and two in MSPs, \cite{dgs16} obtained a $P$--$P_{\rm \mu}$ relation as $P_{\rm \mu}\simeq 1.06P^{0.96}$. This predicts a micro-structure periodicity of 16.2, 15.8 and 4.2\,$\mu$s for the periods of J1022+1001, J2145$-$0750 and J1744$-1134$, respectively, consistent with our measurements as summarised in Table~\ref{tab:obs}. Similarly, from \cite{kjv02}, the micro-structure width (in $\mu$s) scales with the pulsar rotational period (in ms) as $\tau_{\rm \mu}\simeq 0.3 P^{1.1}$. For J1022+1001, J2145$-$0750 and J1744$-1134$, this gives a micro-structure width of 6.5, 6.4 and 1.4\,$\mu$s, respectively, which are qualitatively similar to the measurements in this work. Therefore, our results provide additional support for the extension of the $P$--$P_{\rm \mu}$ and also likely the $P$--$\tau_{\rm \mu}$ relations from the canonical pulsar population to MSPs. 



\subsection{Possible link to fast radio bursts}
\begin{figure}
\centering
\includegraphics[scale=0.4]{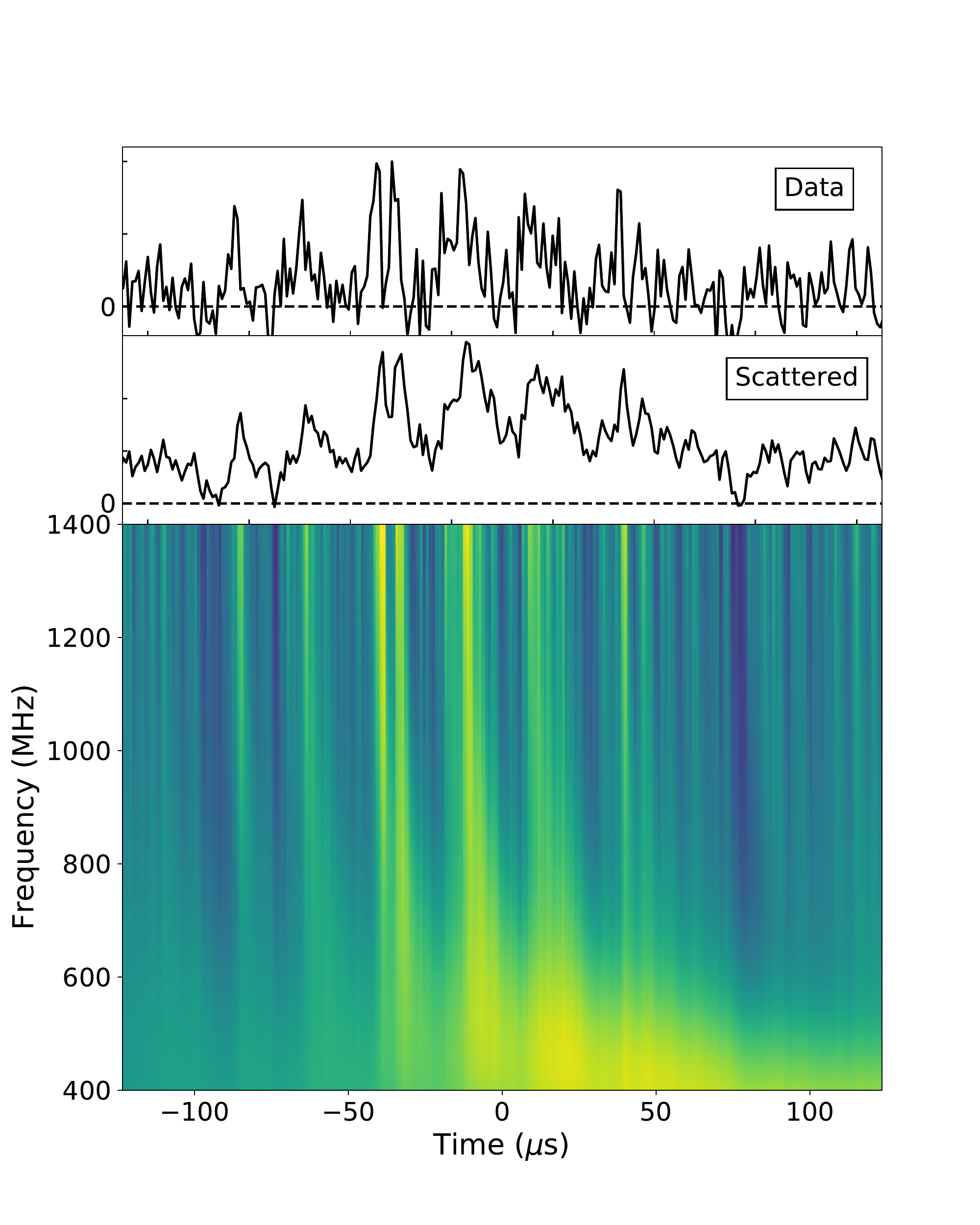}
\caption{Simulation of scattered micro-structure pulse, using the detection from PSR~J1022+1001 shown in the right panel of Figure~\ref{fig:micropulses}. Here we assumed a scattering timescale of 0.6\,$\mu$s at 1.4\,GHz, scattering law of $\tau_{\rm d}\propto f^{-4.4}$ and a spectral index of $-1.8$. \label{fig:scatter1022pulse}}
\end{figure}

Recently, a series of studies have reported the discovery of fast radio bursts (FRBs) which exhibit some stunning narrow temporal emission features, with a timescale down to microsecond or even one-hundred nanosecond level \citep[e.g.,][]{nhk+21a,nhk+21b}. In particular, some show quasi-periodic intensity modulation from a single burst detection \citep{abb+21,nhk+21b,mpp+21}, similar to the micro-structure pulses detected in radio pulsars. Indeed, \cite{abb+21} presented an extensive discussion on the possible origin of quasi-periodic emission in three of the FRBs detected with the Canadian Hydrogen Intensity Mapping Experiment (CHIME) Telescope, where micro-structure emission mechanisms alike those invoked for radio pulsars and magnetars were considered as one of the possible explanations. Two of the FRBs reported there, i.e., FRB 20191221A and FRB 20210206A, appear to show evidence of scattering which creates an apparent overlap between each individual components. While scattering in our observations at L-band is expected to be negligible compared with the time resolution of the data \citep{cl02}, we could still attempt to explore if micro-structure pulses, in case of prominent scattering, can reproduce pattern similar to those seen in the CHIME bursts. In order to do so, we chose one of the pulses from PSR~J1022+1001 shown in Figure~\ref{fig:micropulses} (on the right), and convolved with a pulse broadening function based on the thin screen scattering model \citep{wil72}. We applied a scattering timescale as a function of frequency as $\tau_{\rm d}\propto f^{-4.4}$, assuming a Kolmogorov spectrum, a typical spectral index of $-1.8$ and a $\tau_{\rm d}=0.6$\,$\mu$s at 1.4\,GHz. Figure~\ref{fig:scatter1022pulse} shows the simulated pulse profile starting from our observing frequency, 1.4\,GHz, and spreading down to 400\,MHz, the lower bound of the CHIME band. The simulation managed to reproduce the general feature of the two FRBs reported in \cite{abb+21} (in particular FRB 20210206A in their Figure 1b), where the scattering occurs mostly at the bottom of the frequency band and creates an emission floor that lifts each individual components. It is also important to point out that the original micro-structure pulse from our observation does not have an apparent emission floor a prior to being scattered, which is consistent with what \cite{abb+21} found for the three bursts after de-scattering them. 

Though the first FRBs were found to have an apparently flat P.A. swing \citep{pbj+16}, recent observations have started to also reveal those of more complex P.A. structures \citep[e.g.,][]{lwm+20}. \cite{nhk+21a} showed that some of the bursts from the repeating FRB 20180916B exhibit P.A. curves with a generally flat shape but small variations within about a 10\,deg range. The P.A. of FRB 20210213A also shows similar characteristics \citep{abb+21}. These are compatible with the P.A. curves seen from the micro-structure pulses as shown in Figure~\ref{fig:1022PA}. If FRB emission
comes from neutrons star magnetospheres, the generally flat P.A. of these FRBs may be explained by the fact that the emission is always from a very narrow pulse phase window, i.e., a small fraction of the  magnetosphere where intrinsically the P.A. variation is very limited.
  
Assuming these FRBs with periodic structure can be associated to micro-structure emission from neutron stars, it is then indeed possible to use the $P$--$P_{\rm \mu}$ and $P$--$\delta_{\rm \mu}$ relationships to infer the rotational period of the underlying neutron star. This may provide useful input for the ongoing efforts of searching for potential periodicities in FRBs. But further evidence is needed to make such a link, so that 
we will expand on this idea in a separate forthcoming work.



\section{Conclusions} \label{sec:conclu}
We have detected micro-structure emission in three MSPs, PSRs~J1022+1001, J2145$-$0750 and J1744$-$1134, using highly sensitive observations with LEAP at 1.4\,GHz. A fraction of the pulses show quasi-periodic micro-structure features. In PSR~J1022+1001 and J1744$-$1134 they were seen to be significantly polarised. The occurrence rate of quasi-periodic micro-structure was found to be consistent among pulses with different peak flux densities including the brightest group. Using an ACF analysis we have measured the periodicity and width of the micro-structures in these three pulsars. For PSR~J1022+1001, we showed that the P.A. obtained from micro-structure pulses are from a narrow phase window and on average consistent with that of the average profile, with a small degree of variations. The results have allowed us to further examine the frequency and rotational period dependency of micro-structure properties, and thus the angular beaming and temporal modulation models that explain the appearance of micro-structure. These results have also implied a possible link to FRBs which exhibit a similar emission morphology. 

\section*{Acknowledgements}
We thank Laura Spitler for providing valuable comments on this manuscript. K.~Liu acknowledges the financial support by the European Research Council for the ERC Synergy Grant BlackHoleCam under contract no. 610058. J. Antoniadis is supported by the Stavros Niarchos Foundation (SNF) and the Hellenic Foundation for Research and Innovation (H.F.R.I.) under the 2nd Call of ``Science and Society'' Action Always strive for excellence -- ``Theodoros Papazoglou’' (Project Number: 01431). J. W. McKee is a CITA Postdoctoral Fellow: This work was supported by Ontario Research Fund—research Excellence Program (ORF-RE) and the Natural Sciences and Engineering Research Council of Canada (NSERC) [funding reference CRD 523638-18].

This work was supported by the ERC Advanced Grant ``LEAP", Grant Agreement Number 227947 (PI M.~Kramer). The Effelsberg 100-m telescope is operated by the Max-Planck-Institut f{\"u}r Radioastronomie. Pulsar research at the Jodrell Bank Centre for Astrophysics and the observations using the Lovell Telescope are supported by a consolidated grant from the STFC in the UK. The Westerbork Synthesis Radio Telescope is operated by the Netherlands Foundation for Radio Astronomy, ASTRON, with support from NWO. The Nan{\c c}ay Radio Observatory is operated by the Paris Observatory, associated with the French Centre National de la Recherche Scientifique. The Sardinia Radio Telescope (SRT) is funded by the Department of Universities and Research (MIUR), the Italian Space Agency (ASI), and the Autonomous Region of Sardinia (RAS), and is operated as a National Facility by the National Institute for Astrophysics (INAF). This research is a result of the Europe-wide effort to directly detect gravitational waves using pulsar timing, known as the European Pulsar Timing Array (EPTA). WZ is supported by the the National SKA Program of China (No. 2020SKA0120100) and National Natural Science Foundation of China (No. 12041304,11873067).

\section*{Data Availability}

The timing data used in this article shall be shared on reasonable request to the corresponding author.

\bibliographystyle{mnras}
\bibliography{journals,psrrefs,modrefs,crossrefs}
\end{document}